\documentclass[showpacs,aps,prb,amsmath,amssymb,twocolumn]{revtex4}
\usepackage{graphicx}
\usepackage{dcolumn}
\usepackage{bm}
\usepackage{amsmath,epsfig}
\newcommand{\ket}[1]{\left\vert#1\right\rangle}

\newcommand{\s}{\uparrow}
\newcommand{\g}{\downarrow}
\newcommand{\ug}{\!=\!}
\newcommand{\eq}{Eq.~}
\newcommand{\fig}{Fig.~}
\newcommand{\figs}{Figs.~}
\newcommand{\cfr} {cfr.~}
\newcommand{\ie} {i.e.~}

\begin{document}
\author{Francesco Ciccarello\mbox{$^{1}$}}
\author{Mauro Paternostro\mbox{$^{2}$}}
\author{G. Massimo Palma\mbox{$^{3}$}}
\author{Michelangelo Zarcone\mbox{$^{1}$}}
\affiliation{ \mbox{$^{1}$} CNISM and Dipartimento di Fisica e
Tecnologie Relative, Universit\`{a} degli Studi di Palermo, Viale
delle
Scienze, Edificio 18, I-90128 Palermo, Italy \\
\mbox{${\ }^{2}$} School of Mathematics and Physics, Queen's
University, Belfast BT7 1NN, United Kingdom\\
\mbox{${\ }^{3}$} NEST- INFM (CNR) \& Dipartimento di Scienze
Fisiche ed Astronomiche, Universit\`{a}
degli Studi di Palermo, Via Archirafi 36, I-90123 Palermo, Italy}

\begin{abstract}

We investigate the time evolution of entanglement in a process where
a mobile particle is scattered by static spins. We show that entanglement increases monotonically during a transient and then saturates to a steady-state value. For a quasi-monochromatic mobile particle, the transient time depends only on the group-velocity and width of the incoming wavepacket and is insensitive to the interaction strength and spin-number of the scattering particles. These features do not depend on the interaction model and can be seen in various physical settings.

 \end{abstract}


\title{Rising time of entanglement between scattering spins}
\maketitle
Scattering, an almost ubiquitous mechanism in physics and a rather well-studied topic, has very recently gained renewed interest within the community working on quantum mechanics in virtue of the potential that it has in many respects~\cite{dechiara,imps,ciccarello,ciccarello1}. Scattering processes between two subsystems are effective in order to probe correlation properties of quantum many-body systems~\cite{dechiara}. Moreover, under proper conditions, scattering can be exploited to prepare non-classical states of unaccessible systems~\cite{imps,ciccarello,ciccarello1}. Interesting proposals have been put forward for the generation of quantum correlated light-matter states via off-resonant coherent scattering~\cite{kupryianov}. Frequently, a stationary approach to scattering processes is chosen, especially for the sake of entanglement generation: the system under scrutiny is observed at long-time scales, when it should have reached steady conditions. This cuts the (often complicated) time-evolution from the effective description of scattering dynamics. While this approach is computationally convenient and frequently useful, it is not fully satisfactory since it leaves some interesting questions, related to the details of the dynamical evolution, unanswered~\cite{vanhove}. In particular, it does not give information on the time needed by entanglement to reach its stationary value, which is a pivotal point for the aims of coherent quantum information processing (QIP). In fact, an estimate of such transient time will help us to anticipate, quantify and eventually counteract the effects that decoherence might have in a given process.

In this paper, we focus on a prototypal setting involving both mobile and static spin particles~\cite{imps,ciccarello,ciccarello1} to shed some light onto these issues. We clearly identify the physical parameters that determine the duration of the scattering process. {\it Independently of the Hamiltonian chosen in order to model the spin-spin coupling}, we show that when a quasi-monochromatic incoming wavepacket of the mobile spin is prepared, {with a given average momentum $k_0$}, stationary conditions are reached in a time dictated only by the wavepacket width in momentum space  $\Delta{k}$. This parameter can thus be used to tune the duration of the scattering event and the rate of entanglement generation in the system. Counter-intuitively, the interaction strength of a given spin-spin Hamiltonian model coupling mobile and static particles does not affect the scattering transient time $\Delta{\tau}$ but only determines the maximum entanglement attainable in the process. Although, as anticipated, our conclusions do not depend on the details of the model being considered, most of our quantitative results  are presented for spin-spin interactions induced by the Heisenberg exchange  coupling. This model arises naturally in the spin-spin interaction of magnetic impurities embedded in a one-dimensional  electron nanowire as well as in other situations, including the interaction between a free and a bound electron in a carbon nanotube~\cite{imps}. However, in the second part of our work we quantitatively address the results for {an anisotropic $\text{XYZ}$ model}. Tuning the ratio of its parameters, we are able to span a wide range of significant spin-spin Hamiltonians. 
%Both can be engineered in many physical contexts ranging from solid-state physics to condensed matter and quantum optical settings.  
Remarkably, our results can be extended to an arbitrary number of static particles and intrinsic spin numbers, which makes them valid in a heterogeneous set of physical settings (ranging from spintronics to cavity-quantum electrodynamics)~\cite{imps,ciccarello,ciccarello1,kupryianov,photonmodel}.  In passing, we also reveal an unexpected and general monotonic time-behavior of entanglement, observable under easily-matched conditions. Although wiggling of the reflected wavefunction of the mobile particle is observed due to interference at the scattering site(s), entanglement can only grow in time. First, we empirically observe this behaviour for relevant coupling models and then give a clear physical explanation for the case of the Heisenberg exchange coupling and one scattering center.
The features described above make it clear that a scattering-based mechanism holds the promises for a genuinely control-limited distribution of entanglement in a partially addressable quantum register. In contrast to procedures based on the temporal gating of spin-spin interactions the process addressed in this paper allows for a dramatic relaxation of the time control, which is a major advantage in light of a potential experimental implementation. 

The remainder of the paper is organized as follows. In Sec.~\ref{SecI} we introduce the single-scatterer version of the system addressed in our work and study the entanglement rising-time. An important bench-mark is set by comparing the results of this study with those corresponding to the scattering-less case of two exchange-coupled static particles. Sec.~\ref{SecII} extends the analysis to the two-scatterer case. In Sec.~\ref{SecIII} we quantitatively prove the insensitivity of the entanglement rising-time to the spin-spins coupling strength. Finally, in Sec.~\ref{conclusioni} we summarize our findings and remark their most important implications. 

\section{Scattering by a single static particle}
\label{SecI}

In order to provide a significant mile-stone for our main results, we first review the well-known case of two {\em static} spin-1/2 particles, $e$ and $1$, interacting via the Heisenberg coupling $\hat{H}\!=\!\mathcal{J} \,\hat{\mbox{\boldmath$\sigma$}}\!\cdot\!\hat{\mathbf{S}}_1$. Here, $\hat{\mbox{\boldmath$\sigma$}}$ ($\hat{\mathbf{S}}_1$) is the spin operator of particle $e$ ($1$) and $\mathcal{J}$ is the coupling rate (we use units such that $\hbar\!=\!1$ throughout the paper). We assume the initial state 
\begin{equation} \label{chi0}
\ket{\chi(0)}=\ket{\s\g}_{e1}=\frac{1}{\sqrt{2}}(\ket{\Psi^+}\!+\!\ket{\Psi^-})_{e1},
\end{equation}
where $\ket{\uparrow}$, $\ket{\downarrow}$ are the spin states of each particle and  $\ket{\Psi^{\pm}}_{e1}\!=\!(\ket{\s\g}\pm\ket{\g\s})_{e1}/\sqrt{2}$ are steady states of $\hat{H}$ with energies $\mathcal{J}/4$ and $-3\mathcal{J}/4$, respectively. Upon evolution, the $e\!-\!1$ state reads 
\begin{equation}
\ket{\chi(\tau)}=\frac{1}{\sqrt{2}}(e^{-i\mathcal{J}\tau/4}\ket{\Psi^+}\!+\!e^{i3\mathcal{J}\tau/4}\ket{\Psi^-})_{e1}.
\end{equation}
As a measure of the entanglement between the particles in $\rho_{e1}(\tau)\ug \left\vert \chi(\tau)\rangle\langle\chi(\tau)\right\vert$, we use the logarithmic negativity $E_N(\tau)$~\cite{logneg}. A straightforward calculation gives  $E_N(\tau)\ug \log_2 (1\!+\!|\sin{\mathcal{J}\tau}|)$, which oscillates  with characteristic time $\mathcal{J}^{-1}$.
\begin{figure}[b]
\centerline{\includegraphics[height=0.18\textheight]{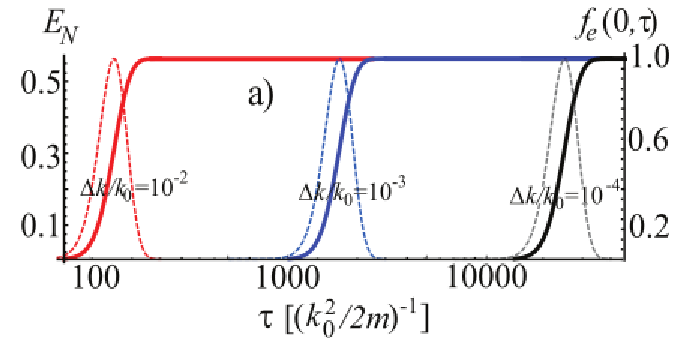}}
\centerline{\includegraphics[height=0.22\textheight]{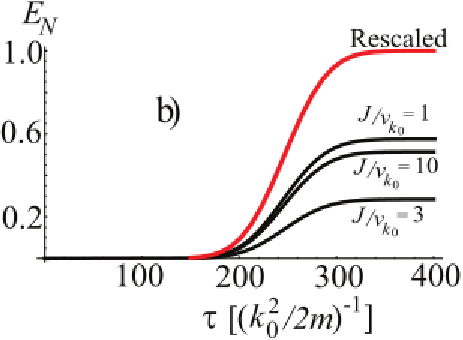}}
\caption{(Color online) 
{\bf a)} $E_N(\tau)$ (full curves) and \emph{ free-propagating}-particle $f_{e}(0,\tau)$ (dashed lines, rescaled to the respective maximum) in a log-lin scale for $x_0\ug 5\Delta x$, $J/v_{k_0}\ug1$ and $\Delta k/k_0\ug10^{-4},10^{-3}$ and $10^{-2}$. {\bf b)} $E_N(\tau)$ for $x_0\ug 5\Delta x$, $\Delta k/k_0\ug10^{-2}$ and $J/v_{k_0}\ug1,3,10$. The top line shows the same curves rescaled to their maximum. The choice of the parameters used in these plots optimize their visibility (other values are equally valid).} 
\label{Fig2}
\end{figure}  
$E_N(\tau)$ is maximized at $\tau\ug (2q\!+\!1)\pi/2\mathcal{J} ^{-1}$ ($q\in{\mathbb N}$), which shows that for a set coupling strength $\mathcal{J}$ a fine control on $\tau$ is required.  This is not the case when the spin-spin interaction takes place during scattering. 

To show it, let us address the case of a mobile particle $e$, interacting with the static spin $1$ during scattering. To fix the ideas, we  consider a one-dimensional wire along the $x$-axis where spin $1$ is embedded at position $x=0$. Spin-spin interaction occurs, via a Heisenberg model, when the mobile spin reaches $x\ug 0$ and is scattered by $1$. Assuming a quadratic dispersion law for $e$, the Hamiltonian reads 
\begin{equation} 
\label{H_1imp}
\hat{H}=\frac{\hat{p}^{2}}{2m^*} + J \hat{\bm \sigma}\cdot
\hat{\mathbf{S}}_1\,\delta(x),
\end{equation}
where $\hat{p}$ ($\,m^*$) is the momentum operator (effective mass) of $e$ and $J$ is the coupling strength (notice that, unlike $\mathcal{J}$, $J$ has the dimensions of a frequency times length). 
The incoming mobile spin $e$ is supposed to have wavevector $\eta k$ ($\eta\ug\pm 1$, $k\!\ge\! 0$) and be prepared in the spin state $\ket{m'_e\ug \,\g,\s}_e$ while the static spin $1$ is initially in $\ket{m'_1 \ug \, \g,\s}_{1}$. We call $\mu\ug\{m'_e,m'_1\}$ ($\nu\!=\!\{m_e,m_1\}$) the set of spin quantum numbers of mobile and static particles before (after) the scattering event. Correspondingly, $r_{k,\eta,\nu}^{\mu}$ ($t_{k,\eta,\nu}^{\mu}$) is the probability amplitude that $e$ is reflected (transmitted) in state $\ket{m_e \ug \, \g, \s}_e$, while the spin-state of $1$ is changed into $\ket{m_1 \ug \, \g,\s}_{1}$. 
These probability amplitudes {depend on $J/v_k$ ($v_k\ug k/m^*$)} and are computed by solving the
time-independent Schr\"{o}dinger equation (SE) associated with Eq.~(\ref{H_1imp}) and imposing proper boundary contitions at $x\ug 0$~\cite{marsiglio}. The steady state of the system $|\Psi_{k,\eta}^{\mu}\rangle$ has wavefunction $\Psi_{k,\eta}^{\mu}(x,\nu)\ug\langle x,\nu|\Psi_{k,\eta}^{\mu}\rangle$ of the form
\begin{equation}
\label{stat_states}
\Psi_{k,\eta}^{\mu}(x,\nu)\ug{e^{i \eta k x}}[(\delta_{\nu,\mu}\!+\!r_{k,\eta,\nu}^{\mu}\,e^{\!-2\!i \eta k x})\theta(-\eta x)
\!+\!\,t_{k,\eta,\nu}^{\mu}\,\theta(\eta x)],\,\,
\end{equation}
where we have omitted a factor $1\!/\!\sqrt{2\pi}$ and $\theta(x)$ is the Heaviside step function. As the kinetic energy of $e$ is the only free-energy term in \eq (\ref{H_1imp}) the system's spectrum is ${\varepsilon}_k\ug k^2/(2m^*)$ and thus coincides with that of a free-propagating $e$. In order to investigate the dynamics of entanglement during scattering, {we consider} $e$ as prepared in a Gaussian wavepacket $\ket{\varphi}$ such that 
\begin{equation}
\langle x|\varphi \rangle=\varphi(x)=(2\pi\beta)^{-1/4}e^{ik_{0}(x\!+\!x_{0})}
e^{-(x\!+x_{0})^{2}/4\beta} 
\end{equation}
 with $x_0,k_0>0$ the average position  and momentum of $e$. As for the uncertainties, we have $\Delta x\ug(\Delta{k})^{-1}\ug\sqrt{2\beta}$.
The overall initial state is taken as  
$\ket{\Psi(0)}\ug\ket{\varphi}\ket{\s,\g}_{e,1}$, whose time-evolved version $\ket{\Psi(\tau)}$ is found by solving the respective time-dependent SE. 

In~\fig\ref{Fig2}, we study the evolution of spin-spin entanglement between $e$ and $1$ when the former is prepared in a quasi-monochromatic wavepacket. Clearly, in the processes at hand, the entanglement associated with the full quantum state involves both motional and spin degrees of freedom. The study of this hybrid form of entanglement, which is in general a difficult task requiring {\it ad hoc} tools for his approach, is beyond the scopes of our work. A possible way to tackle it would be the investigation of the non-locality content of the state of particles $e$ and $1$ and involving both internal and external degrees of freedom, along the lines as in Ref.~\cite{wodkiewicz}. In~\fig\ref{Fig2}(a) we set $J/v_{k_0}=1$, $x_0\ug 5\Delta x$ and plot $E_N(\tau)$ for increasing values of $\Delta k/k_0$ up to 10$^{-2}$. The differences with respect two static spins are striking. First, there is no oscillatory behavior. At small $\tau$, no entanglement is found as far as $e$ has not yet reached $1$. At large times, $e$ is far from spin 1 and $E_N(\tau)$ takes a steady value. At intermediate times, the entanglement shows a {monotonic increase} before eventually saturating to a steady value, which does not depend on $\Delta k$. Second, unlike the case of static spins, the characteristic time $\Delta{\tau}$ of the entanglement evolution is now \emph{ independent} of the spin-spin coupling strength $J$. To illustrate this, in~\fig\ref{Fig2}(b) we set $\Delta k$ and study $E_{N}(\tau)$ for various $J/v_{k_0}$'s. Remarkably, although the steady value of $E_N(\tau)$ depends on $J/v_{k_0}$, the rising time $\Delta \tau$ does not, as it is revealed by the top-most curve in~\fig\ref{Fig2}(b). There, for an assigned value of $J/v_{k_0}$, we have rescaled each $E_{N}(\tau)$ to the respective stationary value and found that the curves are identical. This marks a profound difference with the static-particle case: the introduction of motional degrees of freedom in our problem of interacting spins does not result in a mere spoiling effect of the system spin coherences, as it might be naively expected, but deeply affects the dynamics of the particles involved in the process. To identify the parameters upon which $\Delta \tau$ depends in the dynamical-scattering process, we first observe that in ~\fig\ref{Fig2}(a) an increase in $\Delta k$ of one order of magnitude makes $\Delta \tau$ ten times smaller, suggesting an inverse proportionality between these quantities. Then, in~\fig\ref{Fig2}(a) (dashed lines) we plot the probability density $f_{e}(x\ug 0,\tau)$ of finding particle $e$ at $x\ug 0$ for $J\ug0$, \emph{\ie} the \textit{free-propagation} case. It can be clearly seen that the time needed by $E_N(\tau)$ to reach its steady value coincides with the time during which $e$ is found at $x\ug0$ with non-negligible probability in the absence of any scattering. Thus, by using the free-particle time-energy uncertainty principle~\cite{cohen} we obtain $\Delta \tau\!\sim\! 1/(v_{k_0}\Delta k)$, which explains \cite{nota-time-energy} the aforementioned inverse proportionality between $\Delta \tau$ and $\Delta k$ for a given $k_0$. In other words, in the scattering case and for a quasi-monochromatic mobile particle the characteristic time over which entanglement changes is decided \textit{only} by the kinetic parameters specifying the incoming wavepacket $\varphi(x)$. As for a given $J/v_{k_0}$ the steady value of $E_N(\tau)$ is insensitive to $\Delta k$ [\cfr \fig1(a)] we conclude that the rising time $\Delta \tau$ can be tuned simply by adjusting $\Delta{k}$, with no effect on the stationary value of entanglement (for a set $J/v_{k_0}$). This result is key to the study of the feasibility of scattering-based QIP protocols~\cite{imps,ciccarello,ciccarello1} and the quest for effective ways to relax the control on a system. We have checked that these features hold even when particle $1$ has spin number $s\!>\!1/2$. We point out that although Fig.~1 might at first glance suggest that entanglement rises linearly with time in general this is not the case. Rather, our findings show quite clearly that such functional behavior is dictated by the shape of the incoming wavepacket of particle $e$ (here assumed to be Gaussian).

\section{Scattering by two static particles}
\label{SecII}

We now consider the situation where the mobile spin $e$ is scattered by two static spin-1/2 {particles}, $1$ and $2$, placed at $x\ug0$ and $x\ug d$, respectively. The Hamiltonian is the same as in Eq.~(\ref{H_1imp}) with the inclusion of the term $J\hat{\mbox{\boldmath$\sigma$}}\!\cdot\!\hat{\mathbf{S}}_2\,\delta(x\!-\!d)$. Each steady state $|\Psi_{k,\eta}^{\mu}\rangle$ ($\mu\ug\{m'_e,m'_1,m'_2\}$) has energy ${\varepsilon}_k\ug k^2/(2m^*)$  and differs from \eq(\ref{stat_states}) for the replacement $\theta(\eta x)\!\leftrightarrow\!\theta[\eta x \!-\!(\eta\!+\!1)d/2 ]$ and the addition of $(A_{k,\eta,\nu}^{\mu}\,e^{i \eta k x}\!+\!B_{k,\eta,\nu}^{\mu}\,e^{-i \eta k x})[\theta(x)\!-\!\theta(x\!-\!d)]$. The coefficients $\gamma_{k,\eta,\nu}^{\mu}$ ($\gamma\ug A,B,r,t$) depend implicitly on $J/v_{k}$ and $kd$. As shown in~\cite{ciccarello,ciccarello1}, once boundary conditions at $x\ug0,d$ are imposed, the steady states can be determined. This {configuration} has recently been proposed as a potential way to set entanglement between remote spins~\cite{imps,ciccarello,ciccarello1} via mediation of $e$ and only mild time-control: It is enough to wait for a time $\tau\!\gg\!\Delta\tau$, so that a steady state is reached. {Clearly, the quantification of $\Delta\tau$ in the case of an incoming wavepacket (instead of a plane wave as in Refs.~\cite{imps,ciccarello,ciccarello1}) is key to estimate the influence of dissipation and decoherence affecting the scattering particles. This would be a pivotal point in view of potential experimental implementations. 
We take the initial state $|\Psi(0)\rangle\ug\ket{\varphi,\s}_e\ket{\g,\g}_{12}$, whose time-evolution $|\Psi(\tau)\rangle$ is computed through the time-dependent SE. We focus on the dynamics of entanglement $E_N[\rho_{12}(\tau)]$ between static particles $1$ and $2$. Here
\begin{equation}
\rho_{12}(\tau)\ug\sum_{m_e}\int dx\,\langle x,m_e|\Psi(\tau)\rangle\!\langle \Psi(\tau)|x,m_e\rangle
\end{equation}
is the $1\!-\!2$ state obtained by tracing over the spatial and spin degrees of freedom of $e$. We call 
\begin{equation}
p_{e}(\Omega,\tau)=\int_{0}^{d}dx f_{e}(x,\tau)
\end{equation}
the probability to find $e$ within the interaction region $\Omega \ug\{x\!\!:0\!\le x\le\!d\}$ at time $\tau$.
\begin{figure}
\centerline{\includegraphics[height=0.22\textheight]{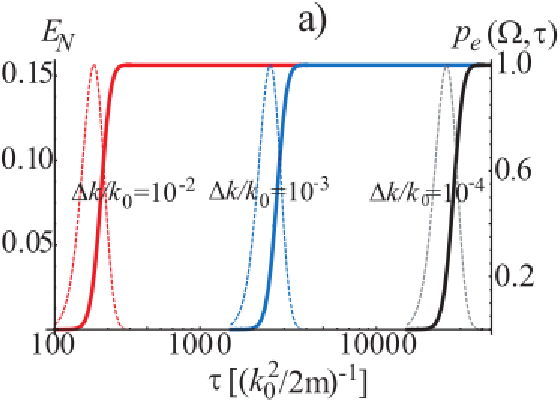}}
\centerline{\includegraphics[height=0.22\textheight]{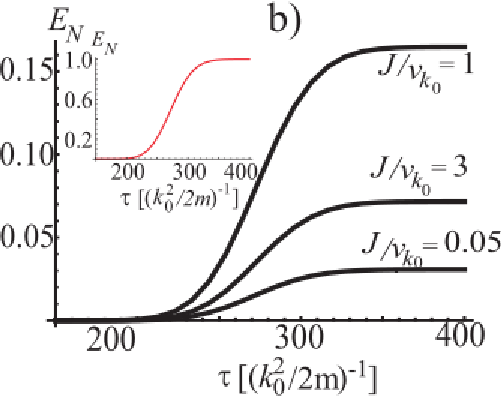}}
\caption{(Color online) {\bf a)} $E_N[\rho_{12}(\tau)]$ for $\Delta k/k_0\ug10^{-2},10^{-3}$ and $10^{-4}$ for $J/v_{k_0}\ug 1$ and $k_0 d\ug\pi$ (solid lines) and \emph{free}-propagating-particle $p_e(\Omega,\tau)$ (dashed lines, rescaled to their maximum). {\bf b)} $E_N[\rho_{12}(t)]$ for $\Delta k/k_0\ug10^{-2}$, $k_0d=\pi$, $J/v_{k_0}\ug 0.5,1$ and $3$. The choice of the coupling strengths used in these plots optimize their visibility (other values are equally valid).}
\label{Fig3}   
\end{figure}
In~\fig\ref{Fig3}(a) we set $x_0\ug 5\Delta x$, $J\!=\!v_{k_0}$, $k_0 d\ug\pi$ and study the behavior of $E_N[\rho_{12}(\tau)]$ and \emph{free}-propagating-particle $p_{e}(\Omega,\tau)$ for increasing values of $\Delta k/k_0$ up to 10$^{-2}$. On the other hand,~\fig\ref{Fig3}(b) considers the effect of different choices of  $J/v_{k_0}$ over the entanglement dynamics, for set values of $\Delta k/k_0$ and $k_0d$. {Features similar to those revealed for a single scattering center are found [see~\figs\ref{Fig2}(a) and (b)]}. Entanglement increases with time to reach a steady value that depends on $J/v_{k_0}$ and $k_0d$. As in the single-scatterer case, $\Delta{\tau}$ is determined only by the time needed by the free-propagating $\varphi(x)$ to cross $\Omega$. In Fig.~\ref{Fig3} our focusing on the regime  $k_0d\ug\pi$ is due to the fact that such setting allows for efficient entanglement distribution  schemes \cite{ciccarello, ciccarello1}. However, we have checked that other values lead to similar conclusions. 

In reaching our conclusions, we have performed a few technical steps which is instructive addressing here. In each of the studied configurations, the time evolution of $\ket{\Psi(0)}\ug\ket{\varphi}\!\ket{\bar{\mu}}$ (with $\ket{\bar{\mu}}$ the total initial spin state~\cite{nota}) can be expanded over the basis of steady states $\{|\Psi_{k,\eta}^{\mu}\rangle\}$ with coefficients given by $\langle \Psi_{k,\eta}^{\mu}|\Psi(0)\rangle$. Provided that $x_0 \!>\! 3 \Delta x$, {\it i.e.} at $\tau\ug0$ spin $e$ is out of the domain $x\!\ge\!0$, 
these scalar products are well-approximated by taking $\theta(-x)\!\simeq\!1$ and {neglecting contributions from other $\theta$-functions} in \eq(\ref{stat_states}), so that
\begin{equation} 
\label{scalar-products}
\langle\Psi_{k,\eta}^{\mu}|\Psi(0)\rangle\!\simeq\! \delta_{{\eta,+}}\delta_{\bar{\mu},\mu}\,\tilde{\varphi}(k)+[\delta_{\eta,+}\,r^{\mu~*}_{k,\eta,\bar{\mu}}\!+\!\delta_{\eta,-}\,t^{\mu~*}_{k,\eta,\bar{\mu}}]\tilde{\varphi}(-k)
\end{equation}
with $\tilde{\varphi}(k')$ the Fourier transform of $\varphi(x)$ (an analogous expression holds in the many-scatterer case). Eq.~(\ref{scalar-products}) allows us to evaluate 
\begin{equation}
\langle x,\mu\ket{\Psi(\tau)}\!\ug\sum_{\mu,\eta}\int_{0}^{\infty} dk \langle \Psi_{k,\eta}^{\mu}|\Psi(0)\rangle e^{-i \varepsilon_k\tau } \Psi_{k,\eta}^{\mu}(x,\mu),
\end{equation}
 which can be solved analytically via a power-series expansion of each spin-dependent amplitude $\gamma_{k,\eta,\nu}^{\mu}$ ($\gamma\ug A,B,r,t$) around the carrier wavevector $k_0$. The replacement of the $n$-th order expansion of these coefficients in $\langle x,\mu\ket{\Psi(t)}$ results in integrals of the form $\int_{0}^{\infty}d\tilde{k}\,e^{-a\tilde{k}^2\!-\!b\tilde{k}}\,\tilde{k}^m$ ($m\ug0,..,n$, $\tilde{k}\ug k/k_0$ and $\mathrm{Re}\,a\!>\!0$), which can be computed in terms of exponential and error functions. The behavior of $f_e(x,\tau)$ is then found by tracing the overall state $\rho(\tau)\ug|\Psi(\tau)\rangle\langle\Psi(\tau)|$ over the spin degrees of freedom. As expected~\cite{cohen}, we find that {during scattering} $f_e(x,\tau)$ wiggles at $x\!\le\!0$ due to interference between incoming and reflected waves. The spin state is found by tracing $\rho(\tau)$ over the spatial and, in the many-scatterer case, spin degrees of freedom of $e$~\cite{mesh}. 

\section{Model-independence of entanglement rising-time}
\label{SecIII}

We are now in a position to explain the insensitivity of $\Delta \tau$ to the values of $J$. {Key to this task is the observation that as long as the wavepacket is narrow enough around $k_0$ the term proportional to $\tilde{\varphi}(-k)$ (with $k\!\ge\!0$) in Eq.~(\ref{scalar-products}) can be neglected. Thus $\langle\Psi_{k,\eta}^{\mu}|\Psi(0)\rangle\!\simeq\!\delta_{{\eta,+}}\delta_{\bar{\mu},\mu}\,\tilde{\varphi}(k)$ meaning that, within the limits of our study, the spectral decomposition of $\ket{\Psi(0)}$ is the one corresponding to $J\ug0$. On the other hand, the spectrum ${\varepsilon}_k$ does not depend on $J$ [see discussion after \eq (\ref{stat_states})] so that the energy and time uncertainties \cite{nota-time-energy} $\Delta{\varepsilon}$ and $\Delta \tau$, are the same as in the free-particle case, \emph{\ie} $\Delta \tau\ug 1/\Delta{\varepsilon}\sim\! 1/(v_{k_0}\Delta k)$. These features hold for a quasi-monochromatic incoming wavepacket.
Most importantly, it is clear that our proof does not rely on the specific form of the interaction Hamiltonian, the number of scattering particles as well as their intrinsic spin numbers. While all these parameters affect the shape of $r^{\mu}_{k,\eta,\bar{\mu}}$'s and $t^{\mu}_{k,\eta,\bar{\mu}}$'s, we have just shown that, for quasi-monochromatic wavepackets, they cannot influence the entanglement rising-time. 

To further illustrate the discussed insensitivity of $\Delta \tau$, in Fig.~3(a) we set $\Delta k/k_0\ug10^{-2}$ and analyze the entanglement between $1$ and $2$ when these are coupled to  $e$ with strengths $J_{e1}\!\neq\!{J_{e2}}$. The case of spin-$1$ scattering centers and equal couplings is also reported.
In Fig.~3(b) we set the same $\Delta k/k_0$ as in Fig.~3(a) and address the XYZ spin-spin model 
\begin{equation}
\hat{H}_{\text{XYZ}}=\sum_{l=x,y,z}J_{l}[\hat{\sigma}_l\hat{S}_{1,l}\delta(x)+\hat{\sigma}_l\hat{S}_{2,l}\delta(x-d)]. 
\end{equation}
We study the cases of $J_x=J_y\ug J_z\!/\!2=v_{k_0}$ (embodying an $\text{XXZ}$ model), $J_x \ug J_y\ug v_{k_0}$ with $J_z\ug 0$  (isotropic $\text{XY}$) and $J_x\ug J_y/2\ug3v_{k_0}$  with $J_z\ug 0$  (anisotropic $\text{XY}$). The ratios of the parameters are chosen so as to provide the best visibility of each plots. The insets in  Figs.~3(a) and (b) report all the curves rescaled to their maximum value, showing that the entanglement rising-time is not affected by the specific quantum spin-number
or the interaction model, which only affect the stationary value of entanglement. In particular, the applicability of our results to the $\text{XY}$ model is remarkable since an effective $\text{XY}$ model is indeed found~\cite{ciccarello1} considering the dispersive interaction of a single photon travelling across a 1D waveguide (a GaAs/GaN nanowire, for instance) and static atom-like systems (such as InAs/GaInN quantum dots or nitrogen-vacancy centers in diamond~\cite{photonmodel}). {The (typical) case  of  linear dispersion law for a photon crossing a waveguide} matches the requirements of our study, \ie a narrow-bandwidth mobile particle. These considerations demonstrate the broad applicability of our investigation and results, which cover a wide range of experimental situations, from spintronics to quantum optics~\cite{imps,ciccarello,ciccarello1,kupryianov,photonmodel}. In the second scenario, in particular, the proven ability to experimentally engineer the temporal shape of photonic wavepackets~\cite{harris} makes scattering-based techniques analogous to the one discussed here and in Refs.~\cite{imps,ciccarello,ciccarello1} quite advantageous against strategies using time-dependent ``modulation" or ``pulsing" of spin-spin interactions. In the former case, once the kinematic properties of the mobile-particle wavepackets are set, there is no necessity for time-control of the entanglement evolution.  We remark that, in order for the steady-state entanglement not to depend on $\Delta k$ [as in~\figs\ref{Fig2} and \ref{Fig3}], this has to be smaller than the inverse of the characteristic length associated with{ $\gamma$-amplitudes}. In particular, for two static spins, it must be $\Delta k\lesssim 1/d$.} 
\begin{figure}[b]
\centerline{\includegraphics[height=0.2\textheight]{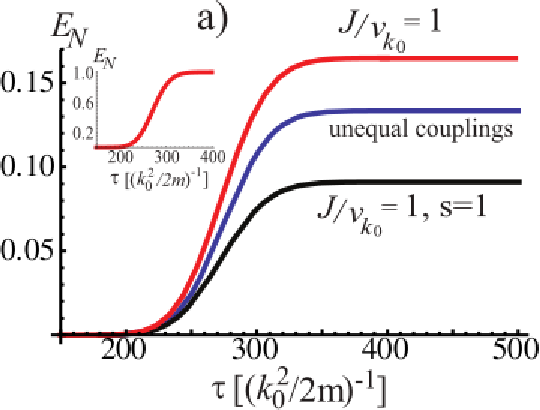}}
\centerline{\includegraphics[height=0.2\textheight]{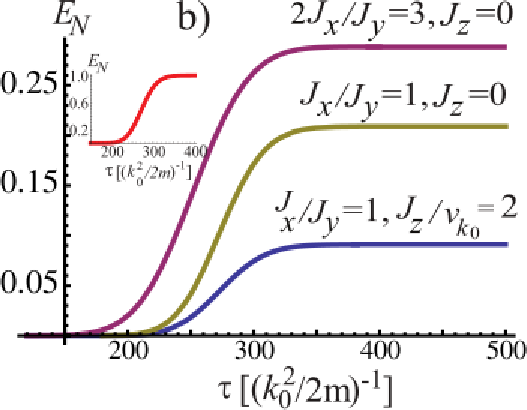}}
\caption{(Color online) {\bf a)} Two-static-spin case, Heinsenberg model. From top to bottom: $E_N[\rho_{12}(t)]$ for $J/v_{k_0}\ug 1$ (equal couplings, spin-1/2), unequal couplings $J_{e1}=2J_{e2}=2.6v_{k_0}$ (spin-1/2) and $J/v_{k_0}\ug 1$ for spin$-1$ scattering centers  (equal couplings). {\bf b)} Two-static-spin case, spin-$1/2$, XYZ model. From top to bottom:  $J_x\ug J_y/2\ug3v_{k_0}$  with $J_z\ug 0$  (anisotropic $\text{XY}$), $J_x\ug  J_y\ug v_{k_0}$ with $J_z\ug 0$  (isotropic $\text{XY}$) and $J_x\ug J_y\ug J_z\!/\!2\ug v_{k_0}$ ($\text{XXZ}$). Insets: curves rescaled to their maximum. We have taken $\Delta k/k_0\ug10^{-2}$ and $k_0 d\ug\pi$. }
\label{Fig4}   
\end{figure}

For the Heisenberg model, the monotonic rise of entanglement~\cite{altroremark} can be interpreted as due to the progressive construction of a phase-difference between spin components of the overall wavefunction. Here, we discuss the steps required in order to prove this feature.  For the sake of argument, we focus on the case of Eq.~(\ref{H_1imp}) and we assume that the incoming wavepacket $\varphi(x)$ is quasi-monochromatic. The Hamiltonian $\hat{H}$ in Eq.~(\ref{H_1imp}) commutes with $\hat{\mathbf{S}}^2$ and $\hat{S}_z$, where $\hat{\mathbf{S}}\ug\hat{\bm{\sigma}}\!+\!\hat{\mathbf{S}}_1$ is the total spin of $e$ and 1. It follows that for the initial spin state $\ket{\Psi(0)}\ug\ket{\varphi}\ket{\s,\g}_{e,1}$ [see also Eq.~(\ref{chi0})]
\begin{equation}
\langle x\ket{\Psi(\tau)}=\frac{1}{\sqrt 2}\sum_{\sigma=\pm}\varphi_\sigma(x,\tau)\ket{\Psi^\sigma}_{e1} 
\end{equation}
%$\langle x\ket{\Psi(t)}\!=\![\varphi_+(x,t)\ket{\Psi^+}\!+\!\varphi_-(x,t)\ket{\Psi^-}]_{e1}/\sqrt{2}$, 
where $\varphi_\pm(x,\tau)$ are evolved wavepackets fulfilling the condition $\varphi_\sigma(x,0)\ug\varphi(x)$ and $\ket{\Psi_{\pm}}_{e1}\ug(\ket{\uparrow\downarrow}\!\pm\!\ket{\downarrow\uparrow})_{e1}/\sqrt{2}$. Upon trace over the spatial variable of $e$ we get $E_N(\tau)\!=\!\log_2(1\!+\!|\mathrm{Im}\Sigma(\tau)|)$ with $\Sigma(\tau)$ the spatial overlap between $\varphi_-(x,\tau)$ and $\varphi_+(x,\tau)$. Now, for $\ket{\Psi^{\pm}}_{e1}$ the spin-spin interaction in Eq.~(\ref{H_1imp}) reduces to an effective static potential $\Gamma_{\pm}\delta(x)$ with $\Gamma_+\ug J/4$ and $\Gamma_- \ug -3J/4$~\cite{marsiglio} with associated reflection and transmission probability amplitudes are 
\begin{equation}
r^{\sigma}_k\ug t^{\sigma}_k-1\ug -\frac{1}{\sqrt{1+(v_{k}/\Gamma_{\sigma})^{2}}}e^{i\mathrm{arccot}(\Gamma_{\sigma}/v_{k})}. 
\end{equation}
The steady states $|\Psi_{k,\eta}^{\pm}\rangle$ are similar to those in Eq.~(\ref{stat_states}) with associated energies ${\varepsilon}_k=k^2/(2m^*)$. Assuming $x_0 \!\geq\! 3 \Delta x$, the projections $\langle\Psi_{k,\eta}^{\pm}|\varphi^{\pm}\rangle$ of $\varphi^{\pm}(x)$ onto the stationary states are well-approximated by a form analogous to Eq.~(\ref{stat_states}) (with due replacements). Provided that the width of the incoming wavepacket around $k_0$ is narrow enough, we find $\langle\Psi_{k,\eta}^{\pm}|\varphi^{\pm}\rangle\simeq\delta_{\eta,+}\,\tilde{\varphi}(k)$. Moreover, $r^{\pm}_k\!\simeq\! r^{\pm}_{k_0}\ug{r}^{\pm}$ [$ t^{\pm}_k\!\simeq\!t^{\pm}_{k_0}\ug{t}^{\pm}$]. Moreover, $r^{\pm}_k\!\simeq\! r^{\pm}_{k_0}\ug{r}^{\pm}$ [$ t^{\pm}_k\!\simeq\!t^{\pm}_{k_0}\ug{t}^{\pm}$]. Thus, by calling $\Delta_r$ ($\Delta_t$) the phase-difference between $r^+$ and $r^-$ ($t^+$ and $t^-$), we get 
%\begin{widetext}
\begin{equation} 
\label{overlap}
\begin{aligned}
\Sigma(\tau)\!&\simeq\!\int_{-\infty}^{0}\!\!dx[|\varphi_{f,R}(x,\tau)|^2\!+\!|r^{+}r^{-}|e^{i\Delta_r }|\varphi_{f,L}(x,\tau)|^2]\\
&+\int_{0}^{\infty}\!\!dx\,|t^{+}t^{-}|e^{i\Delta_t }|\varphi_{f,R}(x,\tau)|^2,
\end{aligned}
\end{equation}
%\end{widetext}
where $\varphi_{f,p}(x,\tau)$ ($p\!=\!R,L$) are evolved {\it free} wavepackets with average positions $x_p\ug{s}_px_0$ at $\tau\ug0$ and Fourier transforms $\tilde{\varphi}(-s_pk')$ with $s_L\!=\!-s_R\!=\!1$. In our derivation
we have neglected terms proportional to $\int_{0}^{\infty}dx\,\varphi^*_{f,R}(x,\tau)\varphi_{f,L}(x,\tau)$ due to the assumption of dealing with an incoming wavepacket narrow enough around $k_0$. {As $\varphi_{f,L}(x,\tau)$ and $\varphi_{f,R}(x,\tau)$ are left- and right-propagating wavepackets, respectively, both $\!\int_{-\infty}^{0}\!\!dx|\varphi_{f,L}(x,\tau)|^2$ and $\!\int_{0}^{\infty}\!\!dx|\varphi_{f,R}(x,\tau)|^2$ are increasing functions of time} whose shapes are similar to those in~Fig.~2(b) for a Gaussian  wavepacket $\varphi(x)$. Hence, the behavior of $|\mathrm{Im}\Sigma(\tau)|$ depends on $\sin{\Delta_{r\,(t)}}$. As $\Delta_{r\,(t)}\!\in\![0,\pi/2]$ (regardless of $J$ and $k_0$) the sine functions are always positive. Therefore, $|\mathrm{Im}\Sigma(\tau)|$ is an increasing function of time, proving the monotonic character of $E_N(\tau)$. Physically, {it is now clear that the build-up of entanglement relies on the phase-difference acquired by the singlet and triplet components of the incoming wavepacket once $e$ is scattered off}. Dynamically, as soon as the scattering process starts and the reflected and transmitted waves are progressively generated, the singlet and triplet components start to build-up the mentioned phase-difference and entanglement grows. For $J/v_{k_0}$ up to $\simeq10$ and $\Delta k/k_0$ up to $10^{-2}$, the largest relative increments of $r^{\pm}_k$ and $t^{\pm}_k$ in the range $[k_0\!-\!3\Delta k,k_0\!+\!3\Delta k]$ are of the order of $\simeq\!10\%$, confirming the validity of our arguments.  Clearly, the monotonic increase of entanglement does not hold for any initial spin state. For instance,  starting from an entangled state of $e$ and $1$, $E_N$ may decrease with time. 

The study of wavepackets having $\Delta{x}\rightarrow 0$, namely the regime opposed to quasi-monochromaticity, is under ongoing investigations.

\section{Conclusions}
\label{conclusioni}

Our work reveals that in a scattering process involving one mobile and many static spins, a situation that can be engineered in various physical settings~\cite{imps,ciccarello,ciccarello1,kupryianov,photonmodel} the time required for entanglement to reach its steady value can be {\it tuned} by preparing the mobile spin's wavepacket, regardless of the specific spin-spin interaction model. If a wavepacket narrow enough in frequency is prepared, the strength of the interaction between mobile and static spins determines just the stationary value of the entanglement which, while scattering takes place, can only grow in time. 
Our findings show that the degree of control on the class of systems we have addressed can be significantly reduced down to the off-line engineering of a proper mobile spin's wavepacket. The quantification of the scattering transient-time enables the analysis of noise effects, such as phase-damping affecting the static spins,  on protocols for entanglement distribution via scattering \cite{imps,ciccarello,ciccarello1}.

\acknowledgments

We thank A. Del Campo, M. S. Kim and D. Sokolovski for discussions. We acknowledge support from PRIN 2006 ``Quantum noise in mesoscopic systems'', CORI 2006, EUROTECH, the
British Council/MIUR through the British-Italian Partnership Programme 2007-2008. MP is supported by the UK EPSRC (EP/G004579/1).

\end{document}